\documentstyle[preprint,aps]{revtex}
\tighten
\begin{document}
\draft
\textheight=8.8 in
\renewcommand{\thesection}{\arabic{section}}
\date{\today}
\preprint{{\Large ${\rm UA/NPPS-97-06}\atop
                   {\rm RUB-TPII-04/97}$}}
\title{Worldline approach to Sudakov-type form factors in non-Abelian
       gauge theories \\
       }
\author{G.\ C.\ Gellas,
        A.\ I.\ Karanikas\footnote{E-mail: akaranik@atlas.uoa.gr},
        and C.\ N.\ Ktorides\footnote{E-mail: cktorid@atlas.uoa.gr}
        }
\address{University of Athens                 \\
         Department of Physics                \\
         Nuclear and Particle Physics Section \\
         Panepistimiopolis                    \\
         GR-15771 Athens, Greece              \\
         }
\author{N.\ G.\ Stefanis\footnote{E-mail:
                                 stefanis@hadron.tp2.ruhr-uni-bochum.de}
        }
\address{Institut f\"ur Theoretische Physik II \\
         Ruhr-Universit\"at Bochum             \\
         D-44780 Bochum, Germany
         }
\maketitle
\newpage
\begin{abstract}
We calculate Sudakov-type form factors for isolated spin-1/2
particles (fermions) entering non-Abelian gauge-field systems.
We consider both the on- and the off-mass-shell case using a
methodology which rests on a worldline casting of field theories.
The simplicity and utility of our approach derives from the fact
that we are in a position to make, {\it a priori}, a more transparent
separation (factorization), with respect to a given scale, between
short- and long-distance physics than diagramatic methods.
\end{abstract}

\newpage   

\section{I\lowercase{ntroduction}}
\label{sec:Intro}
Sudakov effects enter prominently calculations of form factors and
structure functions within Quantum ChromoDynamics (QCD) in connection
with the application of the renormalization group
\cite{CT76,DDT80,Mue81,CS81,Sen81,Kor89}.
In particular, to establish the validity of perturbation theory in
exclusive processes, requires resummation of double-flow Sudakov
logarithms to render factorization sound and isolate them from
single-flow logarithms due to the renormalization group (for a
comprehensive review, see, for example, \cite{CSS89}).

More recently, it was shown that within a modified factorization
scheme, which retains transverse degrees of freedom, the appearance
of Sudakov form factors for bound quarks suppresses large-distance
contributions that invalidate a consistent perturbative calculation
of exclusive processes \cite{BS89,LS92,BOWU95,Ste95,MR97}.
The physical basis for this suppression is that the typical
interquark transverse distance serves as an {\it in situ} infrared
(IR) cutoff for gluon radiation with larger wavelengths.
On the other hand, real and virtual gluons with wavelengths smaller
than these distances but still larger than $1/Q$, $Q$ being the large
momentum transfer involved in the elastic scattering, do not cancel
but accumulate into damping exponentials which give rise to a finite
IR renormalization of the hadron wave functions
\cite{BS89,Ste95,Ste97}.

The derivation of the Sudakov form factor -- originally performed in
\cite{Sud56} for the electromagnetic electron form factor -- within
the diagrammatic context, whether for on- or off-mass-shell fermions,
and whether for an Abelian or a non-Abelian gauge field system,
involves an intricate and quite complex set of arguments that, to a
good extent, divert attention from the physical basis on which they
rest.
Leading contributions to the form factor come from different
integration regions over gluon momenta, each involving physics that
transpires at corresponding ``hard'' and ``soft'' momentum scales and
whose ultimate factorization demands an extremely technical procedure.
For example, the double-flow logarithms derive from integration
regions where soft and collinear boson (gluon) lines overlap, and this
becomes very complicated beyond one loop.
Evidently, a systematic and all-order evaluation in the running
coupling constant of the soft part is required for a hard-scattering
expansion to be useful, and this is by far the most demanding task
involved in any Sudakov-type analysis.

One is naturally led to ask whether these technical features of the
classical diagramatic approaches just mentioned are unavoidable and
due to the physical nature of the problem or whether there exists an
alternative framework in which the main results for Sudakov-type form
factors can be derived in a more transparent and simpler way.
Our point of departure is the formalism we have proposed in a series
of recent works \cite{KKS92,KKS93,KKS95,SKKconf,GKK97a}, where we have
demonstrated the effectiveness of a non-diagrammatic framework for the
computation of closed-form expressions for Green and vertex functions
in the infrared domain.
The theoretical basis of this approach is to recast gauge theories
with spin-1/2 matter fields in terms of path integrals over particle
contours (worldlines) for projecting out their low-energy domain
\cite{Fra66,KS89,JS91,Str92,SS94,KK92}.
It is the purpose of this paper to discuss the derivation of effective
(Sudakov-type) form factors which inhibit the emission of soft gluons
in the framework of the factorization methodology underlying the
aforementioned particle-based approach.

We shall consider both the on- and off-mass-shell case for isolated
spin-1/2 matter particles entering non-Abelian gauge field theories.
In QCD these fermions are hypothetical quarks corresponding to a
physical situation where the quark constituents are so far apart from
each other that each one radiates independently and their Sudakov form
factors resemble those of free fermions.
Abelian systems constitute, of course, special cases and will be
referred to in the end.
The justification and advantages of our approach will become obvious
{\it a posteriori}, given the calculational simplicity of our
treatment and the conceptual clarity by which it attains the called
for separation (factorization) between long- and short-distance
physics.

\section{T\lowercase{hree-point function and fermion elastic
                     scattering}}
\label{sec:3pointfun}
The fermion form factor within the worldline framework is given by
the following connected, three-point function
\begin{eqnarray}
  {\cal G}_{\mu}^{ij}(x,y;z)\,
& = &
  \,\int_{0^{+}}^{\infty} dT
  \int_{\stackrel {\scriptstyle x(0)=x}{x(T)=y}}{\cal D}x(\tau )\,
  \delta\left(x(s)-z\right)
  \int {\cal D}p(\tau ) {\cal P}
  \exp\left\{ - \int_{0}^{s} d\tau
             \left[ i p (\tau ) \cdot \gamma\,+\,m \right]
      \right\}
  \Gamma _{\mu}
\nonumber \\
& &
  \times {\cal P}
  \exp\left\{
             - \int_{s}^{T} d\tau
               \left[ i p (\tau ) \cdot \gamma\,+\,m \right]
      \right\}
  \exp\left[ i \int_{0}^{T} d\tau \, p(\tau )
             \cdot \dot{x}(\tau )\right]
\nonumber \\
& &
  \times \left\langle {\cal P}
  \exp\left[ i g \int_{0}^{T} d\tau \, \dot{x}(\tau ) \cdot
  {\cal A}\left(x(\tau )\right) \right]^{ij}\right\rangle _{A} \; ,
\label{eq:3pGf}
\end{eqnarray}
which involves two fermion lines coupled to a color-singlet current of
the form
$\bar{\psi}(z)\Gamma _\mu\psi (z)$, and where $<\,>_{A}$ denotes
functional averaging of the bracketed quantity in the non-Abelian
gauge field sector, while $\cal P$ denotes path ordering of the
exponential pertaining to the presence of $\gamma$-matrices and/or
non-Abelian vector potentials.

Our basic computational task addresses itself to the expectation value
of a Wilson line operator over paths from an initial point $x$ to a
final point $y$ obliged to pass through the point of interaction $z$
with the external current which injects the large momentum $Q^{2}$.
It is worth pointing out that Wilson line operators facilitate the
factorization process in the diagrammatic approach \cite{CS81,Kor89},
where they make their entrance through operator formalisms and pertain
to lines of {\it semi-infinite} extent.
In contrast, in our case, the Wilson line operators are an integral
part of the worldline casting of the field system and there is no
{\it a priori} reason for which the corresponding line segment cannot
be of {\it finite} extent.
As we shall see in what follows the distinction between Wilson paths
of semi-infinite and such of finite length is of crucial importance in
our scheme.

For a generic non-Abelian system
${\cal A}_{\mu}\equiv A_\mu ^{b}\, t^{b}$
($t^{b}$ being the group generators in the fundamental representation),
a perturbative expansion is called for according to which
\begin{eqnarray}
  \left\langle {\cal P}
  \exp\left[ i g \int_{0}^{T} d\tau \, \dot{x} \cdot {\cal A}
  \right]\right\rangle _{A}
& = &
  1 + (\, i g\, )^2 \Biggl\{ \int_{0}^{s}d\tau _{1}
  \int_{0}^{s} d\tau _{2}\, \theta (\tau _{2} - \tau _{1})
\nonumber \\
& &
  +
  \int_{s}^{T} d\tau _{1} \int_{s}^{T} d\tau _{2}\,
  \theta (\tau _{2}-\tau _{1})
  +
  \int_{0}^{s}d\tau _{1} \int_{s}^{T}d\tau _{2}\Biggr\}
  \dot{x}_\mu (\tau _{1})\dot{x}_\nu (\tau _{2})
\nonumber \\
& &
  \times \left\langle{\cal A}_\mu \left(x(\tau _{1})\right)
  {\cal A}_\nu \left(x(\tau _{2})\right)\right\rangle _A
  + \;{\cal O}(g^4) \; .
\label{eq:poexp}
\end{eqnarray}

In the Feynman gauge and for a dimesionally regularized casting, the
gauge field correlator reads
\begin{equation}
  \left\langle{\cal A}_\mu (x)\,
  {\cal A}_\nu (x')\right\rangle ^{{\rm reg}}_A
\, = \,
  \delta _{\mu \nu}\, C_{F}\, \frac{\mu ^{4-D}}{4\pi ^{D/2}} \,
  \Gamma \left(\frac{D}{2}-1\right)\, |x-x'|^{2-D} \; ,
\label{eq:corr}
\end{equation}
where $C_{F}=(N^{2}-1)/2N$ is the Casimir operator of the
fundamental representation of $SU(N)$.
For notational ease, the indication ``reg'' is omitted in the sequel.

We now isolate what amounts to contributions to Eq.~(\ref{eq:poexp})
from a factorized soft sector of the full theory by staging a
calculation which adopts a straight line path going from $x$ to $z$
(for which we set $\dot{x}(\tau )=u_1$ with $0<\tau <s$) and a second
such path from $z$ to $y$ (for which we set $\dot{x}(\tau )=u_2$ with
$s<\tau <T$).
The no-recoil situation entailed by this restriction, except at point
$z$, suggests that the active gauge-field degrees of freedom entering
our computation are bound by an upper momentum scale which serves to
separate ``soft'' from ``hard'' physics in our factorization scheme.

Setting aside for the moment the issue of a quantitative qualification
of the above remark, let us turn our attention to the cusp occuring at
the point $z$.
It is {\it a priori} obvious that any singular contribution attributed
to what transpires in the immediate vicinity of the cusp is
independent of the geometrical form of the contours connecting $z$
with $x$ and $y$, respectively.
In other words, the specific commitment we have made to straight-line
paths will invariably select these singular contributions, which will
thereby register as multiplicative renormalization constants for the
Wilson line operator.
Such renormalization factors have been originally discussed by
Polyakov \cite{Pol79} and subsequently by other authors
\cite{DV80,Are80,CD81,Aoy82,Bra81} in connection with Wilson
{\it loop} operators.
We shall show, in the framework of the worldline formalism, that the
same situation occurs for {\it open} Wilson lines as well (see also
\cite{CD81,Ste84}).
As it will turn out, the ensuing anomalous dimensions will determine
the renormalization-group evolution of the form factor and eventually,
in the asymptotic limit, will produce Sudakov-type form factors
supported by our worldline factorization scheme.

\section{S\lowercase{oft contribution to the correlator}}
\label{sec:softcorr}
With the above remarks in place, let us proceed with the soft
contribution to Eq.~(\ref{eq:poexp}).
We determine
\begin{eqnarray}
  \left\langle {\cal P}
  \exp\left[ i g \int_{0}^{T} d\tau \, \dot{x} \cdot {\cal A}
  \right]\right\rangle _{A}^{{\rm soft}}
& = &
  1 + (\, i g\, )^2 C_{F}\, \frac{\mu ^{4-D}}{4\pi ^{D/2}} \,
  \Gamma \left(\frac{D}{2}-1\right)
\nonumber \\
& &
   \times \Biggl\{
                  |u_{1}|^{4-D} \int_{0}^{s}d\tau _{1}
                                \int_{0}^{s} d\tau _{2}\,
                  \theta (\tau _{2}- \tau _{1})\,
                  |\tau _{2} - \tau _{1}|^{2-D}
\nonumber \\
& &
   + \, |u_{2}|^{4-D} \int_{0}^{T-s}d\tau _{1}
                      \int_{0}^{T-s} d\tau _{2}\,
                      \theta (\tau _{2}- \tau _{1})\,
                      |\tau _{2} - \tau _{1}|^{2-D}
\nonumber \\
& &
   + \, u_{1} \cdot u_{2} \int_{0}^{s}d\tau _{1}
                          \int_{0}^{T-s} d\tau _{2}\,
        |u_{1}\tau _{1} + u_{2}\tau _{2}|^{2-D}
          \Biggl\} \; + \; {\cal O}(g^{4}) \; .
\label{eq:poexpsoft}
\end{eqnarray}

Now, as long as the parameters $\tau _{i}$\, ($i=1,2$) run over
straight line paths of {\it finite} length, we are dealing with an
off-mass-shell situation, as far as the matter particles are concerned.
Indeed, only for (semi-)infinitely extended paths will their full gauge
field cloud be taken into account.
Notice that such a picture is peculiar to the particle-based, worldline
approach we are currently adopting.
The on-mass-shell situation, corresponding to (straight line) paths of
infinite length, will be taken up later on.

Choosing a frame for which $|u_{1}|=|u_{2}|=|u|$ we find
\begin{equation}
  |u|^{4-D} \int_{0}^{s} d\tau _{1}
            \int_{0}^{s} d\tau _{2}\, \theta (\tau _{2} - \tau _{1})\,
  |\tau _{2} - \tau _{1}|^{2-D}
\,= \,
  - \frac{1}{4-D} \, \frac{1}{D-3} \,
    \left(|u|s\right)^{4-D}
\label{eq:4vel}
\end{equation}
and similarly for the second term inside the curly brackets in
Eq.~(\ref{eq:poexpsoft}), while the third term yields
\begin{eqnarray}
  u_{1} \cdot u_{2} \int_{0}^{s} d\tau _{1}
                    \int_{0}^{T-s} d\tau _{2}\,
                    |u_{1} \tau _{1} + u_{2} \tau _{2}|^{2-D}\,
& = &
  \frac{1}{(4-D)(D-3)}
  \left(|u|s\right)^{4-D}
  F_{D}\left(\frac{s}{T-s},w\right)
\nonumber \\
& &
   + \, \left[|u|\left(T-s\right)^{4-D}\right]
  F_{D}\left(\frac{T-s}{s},w\right) \; ,
\end{eqnarray}
where the relative velocity is given by
$
 w\equiv \frac{u_{1} \cdot u_{2}}{|u_{1}||u_{2}|}
=
 \frac{p_{1} \cdot p_{2}}{|p_{1}||p_{2}|}
$
and
\begin{eqnarray}
  F_{D}(x,w)
& = &
  w\, \Biggl[
             F\left(1,\frac{D-2}{2};\frac{D-1}{2};1-w^{2}\right)
\nonumber \\
& &
   - \, \frac{x+w}{\left(x^{2} + 2 x w + 1\right)^{\frac{D-2}{2}}} \,
 F\left(1,\frac{D-2}{2};\frac{D-1}{2};\frac{1 - w^{2}}{x^{2}
                                              + 2 x w + 1}
  \right)
    \Biggr] \; .
\label{eq:hyper}
\end{eqnarray}

In the limit $D\rightarrow 4_{-}$ we obtain
\begin{eqnarray}
  \left\langle {\cal P}
  \exp\left[ i g \int_{0}^{T} d\tau \, \dot{x} \cdot {\cal A}
  \right]\right\rangle _{A}^{{\rm soft}}
& = &
   1 - \,\frac{g^{2}}{4\pi ^{2}}\, C_{F}\,
         \frac{1}{D-4}\,
         \left[\varphi(w)-2\right]
         \left[ 1 + \frac{D-4}{2}
                    \ln\left(\pi {\rm e}^{2 + \gamma _{E}}\right)
         \right]
\nonumber \\
& &
\mathop{\phantom{1\,}}
   -\, \frac{g^{2}}{4\pi ^{2}}\, C_{F}\,
       \ln (\mu|u|s)
       \left[F_{4}\left(\frac{s}{T-s},w\right) - 1\right]
\nonumber \\
& &
\mathop{\phantom{1\,}}
   -\, \frac{g^{2}}{4\pi ^{2}}\, C_F\,
       \ln (\mu|u|(T-s))
       \left[F_{4}\left(\frac{T-s}{s},w\right) - 1\right]
\nonumber \\
& &
\mathop{\phantom{1\,}}
   -\, \frac{g^{2}}{4\pi ^{2}}\, C_F\,
       \frac{1}{D-4}
       \Biggl[\, F_{D}\left(\frac{s}{T-s},w\right)
\nonumber \\
& &
\mathop{\phantom{1\,-\, \frac{g^{2}}{4\pi ^{2}}\, C_F\,\frac{1}{D-4}\;}}
   +          F_{D}\left(\frac{T-s}{s},w\right) - \varphi (w)
       \Biggr] \; ,
\label{eq:poexpD4}
\end{eqnarray}
where $\gamma _{E}$ is the Euler-Mascheroni constant and
\begin{equation}
  F_{4}(x,w)
=
    \frac{w}{\sqrt{1-w^{2}}} \arctan\frac{\sqrt{1-w^{2}}}{w}
  - \frac{w}{\sqrt{1-w^{2}}}
\arctan\frac{\sqrt{1-w^2}}{x+w} \; ,
\label{eq:F4}
\end{equation}
from which it easily follows that
\begin{equation}
  F_{4}(x,w) + F_{4}\left(\frac{1}{x},w\right)
=
  \frac{w}{\sqrt{1-w^{2}}} \arctan\frac{\sqrt{1 - w^{2}}}{w}
\equiv
  \varphi (w) \; .
\label{eq:phi(w)}
\end{equation}

\section{E\lowercase{ffective low-energy theory}}
\label{sec:efflowth}
This is a good point to pause for some basic assessments.
Given a full microscopic field theory, we have, by utilizing its
worldline casting, isolated a sector of that theory in which the matter
particles are ``dressed'' to a scale that makes them appear extremely
heavy to the active, in this sector, gauge field degrees of freedom.
In particular, the latter are in no position to create
fermion-antifermion pairs nor to derail the spin-1/2 particle entities
from straight-line propagation.
This {\it bona fide} ``soft'' sector  has its own high- and low-energy
domains.
Whereas the second domain presumably coincides with the infrared region
of the full theory, given that they completely overlap, the first one
displays ultraviolet (UV) divergences whose (upper) point of reference,
equivalently the UV cutoff, is the aforementioned separation scale.
To face these divergences we were forced to apply the usual dimensional
regularization techniques.
The emerging singularity structure, reflected in anomalous dimensions
that will be derived shortly, is of a similar nature as the one
prevailing in heavy quark effective theory (see, e.g., \cite{Neu94},
and for some pertinent results \cite{GKK97b}).
In this light, the (arbitrary) mass scale $\mu$ can go as high as the
separation point of the factored out subtheory but can certainly not
exceed that of the momentum transfer involved in the form factor.

Going over to Minkowski space entails the substitution
$
 \frac{w}{\sqrt{1-w^{2}}}\arctan\frac{\sqrt{1-w^{2}}}{w}\,
\rightarrow \,
\newline
\frac{w}{\sqrt{w^{2}-1}}\tanh ^{-1}\left(\frac{\sqrt{w^{2}-1}}{w}\right)
$.
Setting
\begin{equation}
  \Gamma _{{\rm cusp}}
=
  \frac{\alpha _{s}}{\pi}\, C_{F}
  \left[
        \frac{w}{\sqrt{w^{2}-1}}
        \tanh ^{-1}\left(\frac{w^{2}-1}{w}\right) - 1
  \right]
\label{eq:cusp}
\end{equation}
and
\begin{equation}
  \Gamma _{{\rm end}}
=
  - \frac{\alpha _{s}}{2\pi}\, C_{F} \; ,
\label{eq:end}
\end{equation}
for the corresponding anomalous dimensions of the cusp and the
endpoint(s) of the contour, the coefficient $\gamma (\alpha _{s},w)$ of
the divergent term is expressed into the following form
\begin{equation}
  \gamma (\alpha _{s},w)
=
     \Gamma _{{\rm cusp}}(\alpha _{s},w)
  + 2\Gamma _{{\rm end}} (\alpha _{s},w)
=
  \frac{\alpha _{s}}{\pi}\, C_F
  \left[\frac{w}{\sqrt{w^{2}-1}}
  \tanh ^{-1}\left(\frac{\sqrt{w^{2}-1}}{w}\right) - 2
  \right] \; .
\label{eq:cuspend}
\end{equation}
Notice the fact that these anomalous dimensions, which are exclusively
associated with the factored out (soft) sector of the full theory,
display explicit dependence on the momentum.

\section{O\lowercase{ff-mass-shell case}}
\label{sec:offms}
Returning to the vertex function we find, in momentum space,
\begin{eqnarray}
  {\cal G}_{\mu}(p_{1},p_{2})
& = &
  \frac{1}{p_{1} \cdot \gamma - m}\,
  \Gamma _{\mu}\,
  \frac{1}{p_{2} \cdot \gamma - m}
\nonumber \\
& &
    \times
    \left\{
           1 - \frac{1}{D-4}\, \gamma (\alpha _{s},w)
         - \ln\left(\frac{\mu |u|}{\lambda}\right)\gamma (\alpha _{s},w)
         - f(\alpha _{s},w)
    \right\}\,
    +\, {\cal O}(g^{4}) \; ,
\label{eq:vertexmom}
\end{eqnarray}
in which
$\lambda = \frac{|m^{2}-p^{2}|}{m}$
is an off-mass-shellness scale serving as an IR regulator, and
\begin{equation}
  Q^{2}
=
  - \left(p_{1}-p_{2}\right)^{2}
=
  - 2p^{2} + 2p_{1}\cdot p_{2} \; ,
\label{eq:Qscale}
\end{equation}
or, equivalently,
\begin{equation}
  \frac{p_{1}\cdot p_{2}}{|p|^{2}}
=
  w
=
  1 + \frac{Q^{2}}{2|p|^{2}} \; ,
\label{eq:Qw}
\end{equation}
where $p_{1}$ and $p_{2}$ are, respectively, the four-momenta of the
initial and final quark with mass $m$.
Finally, the last term in the curly brackets on the rhs of
Eq.~(\ref{eq:vertexmom}) is given by
\begin{equation}
  f(\alpha _{s},w)
=
  \lim_{D \rightarrow 4_{-}}
  \left[
        2F_{D}(1,w)-\varphi (w)
  \right]\,
        \frac{1}{D-4}\, \frac{\alpha _{s}}{\pi}
\label{eq:fin}
\end{equation}
and is actually finite.

Now, the reference mass scale $\mu$ can range from a minimum value
$\mu _{{\rm min}}$ all the way up to $\mu _{{\rm max}}\sim |Q|$.
Soft contributions transpiring below $\mu _{{\rm min}}$ correspond to
genuine infrared effects which are to be summed over in any physical
expression via the instructions of the KLN theorem \cite{KLN62}.
We consequently demand that for $\mu =\mu _{{\rm min}}$ the form
\begin{equation}
  {\cal G}_\mu ^{(0)}
=
  \frac{1}{p_{1} \cdot \gamma - m}\,
  \Gamma _\mu\,
  \frac{1}{p_{2} \cdot \gamma - m}
\label{eq:freevert}
\end{equation}
of the free vertex function is obtained.
We are thus led to the equation
\begin{equation}
  \ln \left(\mu _{{\rm min}}\, \frac{|u|}{\lambda}\right)
  \gamma (\alpha _{s},w) + f(\alpha _{s},w)
=
  0 \; ,
\label{eq:gammaef}
\end{equation}
which is easily solvable in the limit
$\frac{Q^{2}}{p^{2}}\rightarrow\infty$.
We have, in this case,
\begin{equation}
  \gamma (\alpha _{s},w)
\simeq
  \frac{\alpha _{s}}{\pi}\, C_{F}\, \ln\frac{Q^{2}}{p^{2}}
\hspace{0.1in} , \hspace {0.2in}
  f(\alpha _{s},w)
\simeq
  \frac{\alpha _{s}}{4\pi}\, C_{F}\, \ln ^{2}\frac{Q^{2}}{p^{2}} \; ,
\label{eq:gammafsol}
\end{equation}
whereupon we determine
\begin{equation}
  \mu _{{\rm min}}
=
  \frac{|m^{2}-p^{2}|}{(Q^{2}p^{2})^{1/4}} \; .
\label{eq:mu_min}
\end{equation}
Similar results have been obtained by Ivanov, Korchemsky and Radyushkin
\cite{IKR86,KR86,KR92} (see also \cite{KS84}).

\section{O\lowercase{n-mass-shell case}}
\label{sec:onms}
The on-mass-shell case is reached in the limit where the lengths of
the two paths on either side of the cusp go to infinity while setting,
at the same time, $|u_i|=1,\,i=1,2$.
Finally, the mass $\lambda$ assigned to the gauge fields replaces the
off-mass-shellness as the appropriate scale to provide protection
against infrared divergences.

Taking into account that ($|u|=1$), we obtain
\begin{equation}
  u^{2}\mu ^{4-D}
  \int\frac{d^{D}k}{(2\pi )^{D}}\,
  \frac{1}{k^{2} + \lambda ^{2}}
  \int_{0}^{\infty} d\tau _{1}
  \int_{0}^{\infty} d\tau _{2}\,
  {\rm e}^{-ik \cdot \left(u\tau _{1} - u\tau _{2}\right)}
=
  - \Gamma \left(2 - \frac{D}{2}\right)\,
  \frac{1}{(4\pi )^{D/2}}\,
  \left(\frac{1}{\lambda}\right)^{4-D}
\label{eq:intktau}
\end{equation}
and
\begin{eqnarray}
  u_{1} \cdot u_{2}\,
  \mu ^{4-D}
& \int_{}^{} &
  \frac{d^{D}k}{(2\pi )^{D}}\,
  \frac{1}{k^{2} + \lambda ^{2}}\,
  \int_{0}^{\infty} d\tau _{1}
  \int_{0}^{\infty} d\tau _{2}\,
  {\rm e}^{-ik \cdot \left(u_{1}\tau _{1} + u_{2}\tau _{2}\right)}
\nonumber \\
& &
   =
   2 \Gamma \left(2 - \frac{D}{2}\right)\,
   \frac{1}{(4\pi )^{D/2}}\,
   \left(\frac{\mu}{\lambda}\right)^{4-D}\,
   \frac{w}{\sqrt{1-w^{2}}}
   \arctan \frac{\sqrt{1-w^{2}}}{w} \; ,
\label{eq:intktau12}
\end{eqnarray}
so that the renormalized expression for the vertex function assumes
the form
\begin{equation}
  {\cal G}^{R\,({\rm ms})}_\mu (p_{1},p_{2})
=
  \frac{1}{p_{1} \cdot \gamma - m}\,
  \Gamma _{\mu} \,
  \frac{1}{p_{2} \cdot \gamma -m}\,
  \left[1 - \Gamma _{{\rm cusp}}(\alpha _{s},w)
  \ln\frac{\mu}{\lambda}\right] \; ,
\label{eq:vertexms}
\end{equation}
in which the superscript ms stands for ``on mass shell''.

We observe that the difference between the on- and the off-mass-shell
case, as far as anomalous dimensions are concerned, is the extra
contribution $\Gamma _{{\rm end}}$ to the latter.
On physical grounds this is understood as follows.
Finite worldline segments, corresponding to off-mass-shell situations,
entail sudden accelerations or decelerations of the matter particles
which take place at the open ends.
It is this collinear emission of gauge modes that gives rise to the
additional $\Gamma _{{\rm end}}$ contribution which is present no
matter how minutely off-mass-shell the matter particles may are.

\section{R\lowercase{enormalization group evolution}}
\label{sec:rgevol}
We now turn our attention to the renormalization group equation
which furnishes the evolution of the form factor in the soft sector of
the full theory.
For the on-mass-shell case it reads as follows (S stands for ``soft'')
\begin{equation}
    \left\{\ \mu\frac{\partial}{\partial\mu}
  + \beta (g) \frac{\partial}{\partial g}
  + \Gamma _{{\rm cusp}}(w,g) \right\}
    F_{S}\left(w,\frac{\mu ^{2}}{\lambda ^{2}}\right)
=
  0 \; .
\label{eq:rgeq}
\end{equation}

In the limit $w\rightarrow \infty$ we obtain
\begin{equation}
  \Gamma _{{\rm cusp}}(w,g)
\simeq
  \ln \left(\frac{Q^{2}}{p^{2}}\right)\,
  \Gamma _{{\rm cusp}}(g)
  \hspace{0.1in}, \hspace{0.2in}
  \Gamma _{{\rm cusp}}(g)
=
  \frac{g^{2}}{4\pi ^{2}}\, C_{F} + {\cal O}(g^{4})
\label{eq:gammacusp}
\end{equation}
in accordance with the results found previously in \cite{KR86,Kor89}.

As already mentioned, the above renormalization group scheme operates
strictly within the ``soft'' sector which has been factorized
with respect to the full microscopic theory.
This means that the arbitrary mass $\mu$ runs from a minimum scale,
below which KLN physics takes over, to a maximum scale which serves as
the ``roof'' for the soft sector.
We now set, according to the standard procedure (recall that
$\alpha _{s}=g^{2}/4\pi$ is the running coupling constant),
\begin{equation}
  \alpha (\mu ^{2})
=
  \frac{4\pi}{\beta _{0}\ln (\mu ^{2}/\Lambda ^{2})} \; ,
\label{eq:alphas}
\end{equation}
where
$
 \beta _{0}
=
 \frac{11}{3}C_{A} - \frac{2}{3}n_{F}
$,
$C_{A}=\left(N_{c}+1\right)/2N_{c}$ being the Casimir operator of the
adjoint representation of $SU(3)$, and
$\Lambda \equiv \Lambda _{{\rm QCD}}$ is the characteristic scale of
QCD, entering via dimensional transmutation.

Imposing the boundary condition $F_{S}(w,1)=1$ we obtain in the
leading logarithm approximation, once we have adopted a frame in which
$|p|=|p_{i}|,\,i=1,2$,
\begin{eqnarray}
  F_{S}\left(\frac{Q^{2}}{p^{2}},\frac{\mu ^{2}}{\lambda ^{2}}\right)
& = &
  \exp\left[
            - \ln\left(\frac{Q^{2}}{p^{2}}\right)
              \int_{\lambda ^{2}}^{\mu ^{2}}
              \frac{dt}{2t}\,
              \Gamma _{{\rm cusp}}(g(t))
       \right]
\nonumber \\
& = &
  \exp\left[
           - \frac{C_{F}}{2\pi}\,
             \frac{4\pi}{\beta _{0}}\,
             \ln\left(\frac{Q^{2}}{p^{2}}\right)\,
             \ln\left(\frac{\ln\frac{\mu ^{2}}{\Lambda ^{2}}}
             {\ln\frac{\lambda ^{2}}{\Lambda ^{2}}}
                \right)
       \right] \; ,
\label{eq:sff}
\end{eqnarray}
where $\mu$ now denotes the separation point between the soft and the
hard sectors of the theory.

The fact that the overall form factor $F$, which factorizes into a
``hard'' (H) and a ``soft'' (S) part according to
$F\,=\,F_{{\rm H}}\otimes F_{{\rm S}}$,
should be independent from the separation scale $\mu$, along with the
relation
$
 \frac{\partial}{\partial \ln Q^{2}} \ln F_{{\rm S}}
=
 - \int_{\lambda ^{2}}^{\mu ^{2}}\,\frac{dt}{2t}\,
   \Gamma _{{\rm cusp}}(g(t))
$,
which can be surmised from Eq.~(\ref{eq:sff}), leads to the following
solution in the asymptotic regime
$Q^{2}/\lambda ^{2}\rightarrow \infty$
\begin{equation}
  F\left(Q^{2},\lambda ^{2}\right)
=
  \exp\left\{
            - \frac{C_{F}}{2\pi}\, \frac{4\pi}{\beta _{0}}\,
              \ln\left(\frac{Q^{2}}{\Lambda ^{2}}
                 \right)\,
              \ln\left[\frac{\ln \left(Q^{2}/\Lambda ^{2}\right)}
              {\ln\left(\lambda ^{2}/\Lambda ^{2}\right)}
                 \right]
            + \frac{C_{F}}{2\pi}\, \frac{4\pi}{\beta _{0}}\,
              \ln\left(\frac{Q^{2}}{\lambda ^{2}}\right)
      \right\} \; .
\label{eq:offmsff}
\end{equation}

Similar considerations applied to the off-mass-shell case lead to the
following asymptotic behavior, as
$Q^{2}/M^{2}\rightarrow\infty$,
for the form factor
($M^{2}\equiv -p_{1}^{2}=-p_{2}^{2}$)
\begin{eqnarray}
  F
& \simeq & \,
  \exp\left\{
  - \int_{M^{2}}^{Q^{2}}\frac{dt}{2t}\,
    \ln\left(\frac{Q^{2}}{t}\right)\, \Gamma _{{\rm cusp}}(g(t))
  - \int_{\frac{M^{4}}{Q^{2}}}^{M^{2}} \frac{dt}{2t}\,
    \ln\left(\frac{t\, Q^{2}}{M^{4}}\right)\,
    \Gamma _{{\rm cusp}}(g(t))
      \right\}
\nonumber \\
& = &
    \exp\Biggl\{
    - \frac{C_{F}}{2\pi}\, \frac{4\pi}{\beta _{0}}\,
      \ln\left(\frac{Q^{2}}{\Lambda ^{2}}\right)
      \ln\left[
      \frac{\ln \left(Q^{2}/\Lambda ^{2}\right)}
           {\ln \left(M^{2}/\Lambda ^{2}\right)}
         \right]
\nonumber \\
& &
\mathop{\phantom{\exp\{}}
    + \frac{C_{F}}{2\pi}\, \frac{4\pi}{\beta _{0}}\,
      \ln\left(\frac{Q^{2}\Lambda ^{2}}{M^{4}}\right)\,
      \ln\left[
               \frac{\ln \left(Q^{2} \Lambda ^{2}/M^{4}\right)}
                    {\ln\left(\Lambda ^{2}/M^{2}\right)}
         \right]
        \Biggr\} \; .
\label{eq:onmsff}
\end{eqnarray}
Two basic readjustments, with respect to the on-mass-shell case, are
entailed here.
First, the lower limit of integration is given by the mass scale
defined by Eq.~(\ref{eq:mu_min}), meaning that $\lambda$ in
Eq.~(\ref{eq:sff}) is replaced by $\mu _{{\rm min}}$.
In this connection, let us observe that to incorporate this adjustment
into the diagrammatic analysis, Korchemsky (second paper of
\cite{Kor89}) follows Fishbane and Sullivan \cite{FS71} (see also
\cite{KR86}) to introduce an extra, ``infrared'' component
$F_{{\rm IR}}$ to his factorization formula, in addition to the
``soft'', ``jet'' (collinear subgraphs), and ``hard'' terms.
As already noted, our factorization procedure encompasses all
non-hard contributions into the soft sector of the full theory.
Second, an additional term enters the anomalous dimension which,
however, is not momentum-dependent, (cf. Eq.~(\ref{eq:cuspend}), and
thereby gives a much weaker contribution to the form factor in the
asymptotic limit.

\section{A\lowercase{belian case}}
\label{sec:Abel}
To adjust the above results to the Abelian case, we proceed as follows.
We first observe, focusing on the on-mass-shell case, that
\begin{equation}
  \frac{\ln Q^{2}/\Lambda ^{2}}{\ln\lambda ^{2}/\Lambda ^{2}}
\simeq
  1 + \frac{\beta _{0}}{4\pi}\alpha _{s}(Q^{2})
  \ln\frac{Q^{2}}{\lambda ^{2}} \; .
\label{eq:scales}
\end{equation}
The above relation along with Eq.~(\ref{eq:alphas}), when inserted
into Eq.~(\ref{eq:offmsff}), give, in the limit
$Q^{2}/\lambda ^{2}\rightarrow\infty$,
\begin{eqnarray}
  F
& \simeq &
  \exp\left\{
             \frac{C_{F}}{2\pi}\,
             \left(\frac{4\pi}{\beta _{0}}\right)^{2}\,
             \frac{1}{\alpha _{s}}\,
        \left[  \alpha _{s}\frac{\beta _{0}}{4\pi}
                \ln \frac{Q^{2}}{\lambda ^{2}}
              - \frac{1}{2}\, \alpha _{s}^{2}\,
                \left(\frac{\beta _{0}}{4\pi}\right)^{2}\,
                \ln ^{2} \frac{Q^{2}}{\lambda ^{2}}
        \right]
      - \frac{C_{F}}{2\pi}\, \frac{4\pi}{\beta _{0}}\,
        \ln \frac{Q^{2}}{\lambda ^{2}}
      \right\}
\nonumber \\
& = &
     \exp\left\{
                - \frac{\alpha _{s}}{4\pi}\, C_{F} \,
                  \ln ^{2} \frac{Q^{2}}{\lambda ^{2}}
         \right\} \; ,
\label{eq:ffasyoms}
\end{eqnarray}
which coincides with the well-known result, established in
\cite{CT76,BU80,ES81}.
In reference to QED, the familiar result obtained by Jackiw \cite{Jac68}
(see also \cite{FS71,KAR76,Mue79,Col80})
\begin{equation}
  F_{{\rm ms}} \,
= \, {\rm e}^{- \frac{g^{2}}{16\pi}\,
              \ln ^{2}\frac{Q^{2}}{\lambda ^{2}}
              }
\label{eq:ffQEDms}
\end{equation}
is recovered.
Similar considerations produce the off-mass-shell result
\begin{equation}
  F_{{\rm off-ms}}\,
= \,
  {\rm e}^{- \frac{g^{2}}{8\pi}\,
           \ln^ {2}\left(\frac{Q^{2}}{M^{2}}\right)
           } \; ,
\label{eq:ffQEDoffms}
\end{equation}
which was what Sudakov obtained to begin with \cite{Sud56}.
Note the extra factor of 2 entering the exponent, as compared to
Eq.~(\ref{eq:ffQEDms}); it comes from the combined contribution of the
two terms entering the exponent in Eq.~(\ref{eq:onmsff}) and underlines
the significance of the lower limit $\mu _{{\rm min}}$ associated with
the off-mass-shell case.

\section{C\lowercase{oncluding remarks}}
\label{sec:concl}
In conclusion, the present approach to the derivation of Sudakov-type
form factors for spin-1/2 (isolated) matter fields in a non-Abelian
gauge field theory has avoided to face intricate issues in a
diagramatic context, such as the role of jet subdiagrams and how they
communicate with soft and hard parts, or how the operator formalism
comes to the aid of field theory, etc.
The clean way by which a soft sector of the full theory can be
factorized and, basically, determine the form factor structure is, we
believe, a notable accomplishment of the worldline scheme.
The next step, of course, is to assess the merits of our approach to
exclusive hadronic processes like hadron form factors or hadron-hadron
elastic scattering.

\acknowledgments
One of us (G. C. G) wishes to acknowledge a grant from the Bodosakis
Foundation in support of this work.

\newpage   


\end{document}